\documentclass[onecolumn]{aa} 

\usepackage{multirow}
\usepackage{graphicx}
\usepackage{xcolor} 
\usepackage{txfonts}
\usepackage{amsmath} 
\usepackage{dirtytalk}

\newcommand{\vect}{\overrightarrow}

\newcommand{\vEnp}{\vect{E}_{\mathrm{np}}}
\newcommand{\vEp}{\vect{E}_{\mathrm{p}}}

\newcommand{\Ep}{E_{\mathrm{p}}}

\newcommand{\Enpx}{E_{\mathrm{np},x}}
\newcommand{\Epx}{E_{\mathrm{p},x}}

\newcommand{\Enpy}{E_{\mathrm{np},y}}
\newcommand{\Epy}{E_{\mathrm{p},y}}

\bibpunct{(}{)}{;}{a}{}{,} 

\newcommand{\msgguy}[1]{{\textcolor{black}{#1}}}
\newcommand{\redguy}[1]{{\textcolor{black}{#1}}}

\begin{document}

\title{Biases induced by retardance and diattenuation in the measurements of long-baseline interferometers}



   \author{G. Perrin
          \inst{1}\thanks{\redguy{formerly LESIA, Observatoire de Paris, Universit\'e PSL, CNRS, Sorbonne Universit\'e, Universit\'e Paris Cit\'e, 5 place Jules Janssen, 92195 Meudon, France}}
          }

    \institute{\redguy{LIRA}, Observatoire de Paris, Universit\'e PSL, CNRS, Sorbonne Universit\'e, Universit\'e Paris Cit\'e, 5 place Jules Janssen, 92195 Meudon, France\\
              \email{guy.perrin@obspm.fr}
             }

   \date{Received 19 July 2024; accepted 16 December 2024}

 
  \abstract
   {The coherence of long-baseline interferometers is affected by the polarization properties of the instrument. This is a possible source of biases, which would need to be calibrated.}
   {The goal of this paper is to study the biases due to retardance and diattenuation in long-baseline interferometers. In principle, the results can be applied to both optical and radio interferometers. }
   {We derived theoretical expressions for biases on fringe contrast and fringe visibility phase for interferometers whose polarizing properties can be described by beam rotation, retardance, and diattenuation. The nature of these biases are discussed for natural light, circular and linear polarization, and partially polarized light. Expansions were obtained for small degrees of polarization, small differential retardance, and small diattenuation.}
   {The biases on fringe contrasts were already known. \redguy{It is} shown \redguy{in this paper} that retardance and diattenuation are also sources of bias \redguy{on} the visibility phases and derived quantities. In some cases, the bias is zero (for non-polarizing interferometers with natural or partially circulary polarized light.) If the retardance is achromatic, differential phases are not affected. Closure phases are not affected to the second order for an interferometer with weak diattenuation and weak differential retardance and for moderately polarized sources whatever the type of light.   
   Otherwise, a calibration procedure is  required. It has been shown that astrometric measurements are biased in the general case. The bias depends on both the polarization properties of the interferometer and on the $(u,v)$ sampling. In the extreme case where the samples are aligned on a line crossing the origin of the spatial frequency plane, the bias is undetermined and can be arbitrarily large. In all other cases, it can be calibrated if the polarizing characteristics of the interferometer are known. In the case of a low differential retardance and low degree of polarization, the bias lies on a straight line, crossing the astrometric reference point. If the degree of linear polarization varies during the observations, then the astrometric bias has a \redguy{remarkable} signature, which describes a section of the line. For slightly polarizing interferometers, a fixed offset is added without changing the shape of the bias. }
   {A polarizing interferometer does generate bias on visibility contrast and visibility phase. The bias depends on the polarization characteristics of the source. In any case, the bias can be computed if the polarization characteristics of the interferometer are known. Astrometric biases can also be corrected and depend on the $(u,v)$ sampling achieved for the measurements.}


   \maketitle
%

\section{Introduction}
Long baseline interferometers measure the spatial coherence of light. The degree of coherence measured on a baseline is the normalized scalar product of the waves collected by two telescopes averaged over time. Thus, the polarization characteristics of the waves (i.e., a characteristic of the source or of the interferometer)  play a role in the  obtained result. It is \redguy{a well-known fact} that polarization axes cannot be crossed; otherwise the fringe contrast will be canceled for incoherent linear polarizations or a $\pi$ differential birefringence phase will lead to a zero contrast if polarizations are not split. This was \redguy{experienced} by Labeyrie before the first detection of long-baseline fringes at visible wavelengths \citep{Labeyrie1975}. Futhermore, \citet{Perraut1996}  investigated the effects of differential phases between interference patterns and of differential rotation between polarization planes on fringe contrast applied to the  REcombinateur pour GrAnd INterf\'erom\`etre (REGAIN) beamcombiner of the   Grand Interf\'erom\`etre \`a 2 T\'elescopes (GI2T) interferometer, a descendant of Labeyrie's first interferometer.

Optical interferometers have complex trains of mirrors and include delay lines. Accurately modeling their polarization properties is quite complex. The Jones calculus for fully polarized light and the Mueller calculus for partially polarized light are the usual methods applied to deduce the effect of the interferometer on input light waves. \citet{Elias2001} and \citet{Elias2004}  proposed a general formalism to deal with polarization in interferometers by describing the optical train with Jones and Mueller matrices to propagate polarization states and Stokes parameters across the interferometer. This method has been used by \citet{Widmann2023} to study polarization characteristics of the  Very Large Telescope Interferometer (VLTI) and their effects. This method is accurate, but it does not allow for an analytical study of the effects of diattenuation and retardance on visibility data. I have investigated the possibility of modeling complex trains of mirrors using a simple Jones matrix with generalized neutral axes in \citet{Perrin2024b}. I have shown that the VLTI train can be modeled with a quasi-unitary Jones matrix with a very good accuracy because of the large set of optical elements. This method greatly simplifies the problem and the polarization properties of the optical train can then be described by a rotation, a retardance, and a polarizing transmission applied to two orthogonal linear polarizations. I use this formalism in this paper to derive the biases on fringe contrast (visibility modulus), visibility phase, and astrometric quantities.\\
The interferometric formalism is described in Sect.~\ref{sec:formalism}, the impact of the polarization characteristics on the interferograms is given in Sect.~\ref{sec:impact}. The bias on phases is derived in Sect.~\ref{sec:symmetric} and discussed for visibility phases and derivations in Sect.~\ref{sec:phase_bias}. 
In Sect.~\ref{sec:astrometry}, the astrometric bias induced by retardance and diattenuation is discussed. Our conclusions are presented in Sect.~\ref{sec:conclusion}.

%
%
%
%
 %
%
%
\section{Interferometric formalism}
\label{sec:formalism}
%
%
\subsection{General expression of the waves}
The complex notation is chosen to describe the instantaneous wave $\vect{E}$ in the general case:
\begin{equation}
\vect{E}(\vect{r},t)=e^{-i(\vect{k}.\vect{r}-\omega t)}\left( E_x \, e^{-i\varphi_x}\,\vect{e_x} + E_y \, e^{-i\varphi_y}\,\vect{e_y} \right)
.\end{equation}
Here,  $\vect{k}$ is the wave vector \msgguy{and $\omega$ the pulsation of the wave}. It describes the direction of propagation of the wave and depends on the wavelength. Then, $(x,y)$ defines a plane perpendicular to $\vect{k}$. Based on the assumption that a monochromatic wave is a good description of the wave, it is possible to drop the integration over the wavelength that is necessary to describe a quasi-monochromatic wave, as done in~\citet{Goodman}. The full expressions can be recovered with an integration. In the following, $\vect{k}$ defines an axis whose coordinate is noted $z$. The equation above can be written:
\begin{equation}
\vect{E}(z,t)= e^{-i(kz-\omega t)}\left( E_x\,e^{-i\varphi_x}\,\vect{e_x} + E_y\, e^{-i\varphi_y}\,\vect{e_y} \right)
\label{eq:no_bir}
.\end{equation}
This expression holds as well if the $z$ axis is along a curved waveguide, with the $(x,y)$ plane being locally perpendicular to the $z$ axis. 

Here, $\varphi_x$ and $\varphi_y$ are constant for polarized light: $\varphi_x=\varphi_y$ for linearly polarized light and $\varphi_x=\varphi_y\pm\frac{\pi}{2}$, for circularly  polarized light ($+$ for left-handed circular polarization and $-$ for right-handed). $E_x$ and $E_y$ are also constant for polarized light: they can be written $E_x =E_0\cos \alpha$ and $E_y = E_0\sin \alpha$ where, in the case of linearly polarized light, $\alpha$ is the electric vector position angle (EVPA) and $\alpha = \frac{\pi}{4}$ for circularly polarized light. 
For the general case of elliptically polarized light, $\varphi_x$, $\varphi_y$, $E_x$, and $E_y$ can take any value. In the case of unpolarized light, $E_x\,e^{-i\varphi_x}$ and $E_y\,e^{-i\varphi_y}$ take random values as the direction of polarization is random and as the wave components on the two axes are not coherent, with $<E_x^2>=<E_y^2>=E_0^2/2$. In all cases, the intensity carried by the wave is $I_0=E_0^2$.
%
%
\subsection{Polarization properties of the beam path}
During propagation, the wave may experience various effects such as retardance, which will build phase differences between different polarization axes, or diattenuation with a differential transmission depending on the polarization axis. I also consider beam rotation in addition to these effects in this work. I assume that the whole system can be described as a non-depolarizing system and that two perpendicular neutral axes $x$ and $y$ can be defined with a good approximation, as in \citet{Perrin2024b}. This allows for an analytical study the effects of retardance and diattenuation. Retardance could be the consequence of differential phases at reflections on optical surfaces or through thin layers, or of birefringence because of propagation in birefringent media induced by the dependence of optical index with polarization axes. Equation~(\ref{eq:no_bir}) is then updated taking into account two phases $\psi_x$ and $\psi_y$, whose difference leads to the retardance: 
\begin{equation}
\label{eq:with_bir+}
\vect{E}(z,t)=e^{-i(kz -\omega t)}\left( E_x\,e^{-i(\varphi_x+\psi_x)}\,\vect{e_x} + E_y\,e^{-i(\varphi_y+\psi_y)}\,\vect{e_y} \right)
.\end{equation}
In addition, the optical system may be polarizing with a differential throughput on the $x$ and $y$ axes with transmissions $\tau_x$ and $\tau_y$ leading to a new general expression for the wave:
\begin{equation}
\label{eq:with_bir_tau}
\vect{E}(z,t)=e^{-i(kz -\omega t)}\left( \tau_x E_x\,e^{-i(\varphi_x+\psi_x)}\,\vect{e_x} + \tau_y E_y\,e^{-i(\varphi_y+\psi_y)}\,\vect{e_y} \right)
.\end{equation}
Finally, the beams may be rotated in the interferometer by an angle, $\theta,$ thereby mixing the neutral axes of polarization according to:
\begin{equation}
\label{eq:with_bir_tau_rot}
                \vect{E}(z,t)\!=\!e^{-i(kz -\omega t)}\!\left[\begin{array}{ll}
                      \!\!\!\tau_x \cos\!\theta\,E_x\,e^{-i(\varphi_x+\psi_x)}\! - \!\tau_y \sin\!\theta\,E_y\,e^{-i(\varphi_y+\psi_y)} \\
                      \!\!\!\tau_x \sin\!\theta\,E_x\,e^{-i(\varphi_x+\psi_x)}\! + \!\tau_y \cos\!\theta\,E_y\,e^{-i(\varphi_y+\psi_y)} 
                     \end{array} \!\!\!\right] 
.\end{equation}
This general expression is the one described in~\cite{Perrin2024b}, where the Jones matrix describing the effect of the optical system on polarizations is a quasi-unitary matrix; however, there is no frame rotation of the neutral axes \redguy{here} with respect to the $x$ and $y$ axes.
%
%
\subsection{Case of partially polarized waves}
For partially polarized waves, the non-polarized and polarized components of the instantaneous wave are noted $\vEnp(z,t)$ and $\vEp(z,t)$. The instantaneous components on the $x$ and $y$ axes in Eq.~\ref{eq:no_bir} are noted as $\Enpx$ and $\Enpy$ for the non-polarized wave and $\Epx$ and $\Epy$ for the polarized wave.\\ \\
%
%
The degree of polarization is defined as:
\begin{equation}
P=\frac{I_{\mathrm{p}}}{I_{\mathrm{p}}+I_{\mathrm{np}}}
,\end{equation}
where the average intensities of the beams are expressed as:
\begin{equation}
I_{\bullet \mathrm{p}}=\left< ||\vect{E}_{\bullet \mathrm{p}}(z,t)||^2\right>_t= \tau_x^2 E_{{\bullet \mathrm{p}},x}^2 +\tau_y^2 E_{{\bullet \mathrm{p}},y}^2
,\end{equation}
with the $\bullet \mathrm{p}$ index being either the np index for non-polarized light or the p index for polarized light.
%
For the non-polarized component, $< \Enpx^2> = <\Enpy^2> = I_{\mathrm{np}}/2$ for a 100\% \msgguy{throughput}. For a  polarized wave, $\Epx = \Ep \cos\alpha$ and $\Epy = \Ep \sin\alpha$ where $\alpha$ is the position angle of the linearly polarized wave and $\alpha = \frac{\pi}{4}$ for a circularly polarized wave. In either case, $\Ep^2 = I_{\mathrm{p}}$. 
%
%
\subsection{Interferometric equations}
The interferometric equations are derived for a two-beam interferometer. The beams are labeled  1 and  2. The equations just need to be replicated per baseline for a larger number of beams.  
The interferogram is denoted as $I$ and expressed as:
\begin{equation}
 I =\left<| E^1+E^{2*} |^2\right>=I_1+I_2+2\,\mathrm{Re} \left< E^1E^{2*} \right>
 .\end{equation}
In the following, I assume that the intensities $I_1$ and $I_2$ are equal (it is the case after photometric calibration in single-mode interferometers, an imbalance parameter needs to be introduced otherwise) and  noted as $I_0$. The modulated part of the interferogram is noted $\tilde{I}=\mathrm{Re} \left< E^1E^{2*} \right>$. With these notations, the interferogram is:
\begin{equation}
 I =\left<| E^1+E^{2*} |^2\right>=2\,I_0+2\,\tilde{I}
 .\end{equation}
The Poynting vector of the sum of the waves of two neutral axes is the sum of their respective Poynting vectors. As a consequence, the interferogram of the two orthogonal axes of polarization is the sum of the interferograms on each polarization axis. The polarization axes can therefore be studied independently and summed {a posteriori}. In addition, unpolarized light cannot interfere with polarized light as they are mutually incoherent \redguy{(as a matter of fact, the cross terms  like for example $<\!e^{-i(\varphi_{\mathrm{np},x}-\varphi_{\mathrm{p},x})}\!\!>$ are equal to zero as phase $\varphi_{\mathrm{np},x}$ is a random variable, while $\varphi_{\mathrm{p},x}$ takes a constant value)} so that the polarized and unpolarized interferograms can be summed as well. Such conclusions are not necessarily valid for different polarization states and, in the following,  the polarized part of the radiation is considered to be of one type only: either linear or circular. 
%
Two interferograms, $I_x$ and $I_y$, are defined, one per neutral axis. Each interferogram is the sum of the polarized and of the unpolarized interferograms and consequently:
\begin{equation}
\label{eq:mod}
 \left\{\begin{array}{@{}l@{}}
\tilde{I}_x =\mathrm{Re} \left< E_{\mathrm{p},x}^{1}E_{\mathrm{p},x}^{2*} \right> + \mathrm{Re} \left< E_{\mathrm{np},x}^{1}E_{\mathrm{np},x}^{2*} \right> = \tilde{I}_{\mathrm{p},x}+\tilde{I}_{\mathrm{np},x} \\ 
\tilde{I}_y =\mathrm{Re} \left< E_{\mathrm{p},y}^{1}E_{\mathrm{p},y}^{2*} \right> + \mathrm{Re} \left< E_{\mathrm{np},y}^{1}E_{\mathrm{np},y}^{2*} \right> = \tilde{I}_{\mathrm{p},y}+\tilde{I}_{\mathrm{np},y} 
\end{array}\right.
,\end{equation}
The object is supposed to be a point source, so that no other spatial coherence effect is taken into account in this paper. \msgguy{For the case of extended sources, we refer the remarks at the end of Sect.~\ref{sec:symmetric}.}  
%
%
\section{The impact of the polarization properties of the optical system on the interferograms}
\label{sec:impact}
%
%
We first define a series of phase differences and averages where the prefix A is for an average and $\Delta$ for a difference to ease the expression of the equations:
\begin{equation}
\label{eq:def}
\begin{array}{@{}l@{}l@{}}
\Delta \psi_1 = \psi_{x_1} - \psi_{y_1}        \;\;\;\;\;\;\;\;\;\;\;\;(\mathrm{retardance\,in\,beam\,1}), \\
\Delta \psi_2 = \psi_{x_2} - \psi_{y_2} \;\;\;\;\;\;\;\;\;\;\;\;(\mathrm{retardance\,in\,beam\,2}), \\
\mathrm{A} \psi_1 = \frac{\psi_{x_1} + \psi_{y_1}}{2},\\
\mathrm{A} \psi_2 = \frac{\psi_{x_2} + \psi_{y_2}}{2},\\
\Delta\Delta\psi = \Delta \psi_2 - \Delta \psi_1 \;\;\;\;\;\;\;\;(\mathrm{differential\,retardance}), \\
\mathrm{A} \Delta\psi = \frac{\Delta \psi_1 + \Delta \psi_2}{2} \;\;\;\;\;\;\;\; \;\;\;\; \,(\mathrm{average\,retardance}),\\
\Delta \mathrm{A}\psi = \mathrm{A} \psi_1 - \mathrm{A} \psi_2,\\
\gamma = (\varphi_x - \varphi_y) + \mathrm{A} \Delta\psi. 
\end{array}
\end{equation}
%
The two waves of Eq.~\ref{eq:with_bir_tau_rot} collected by telescopes 1 and 2 are noted as $\vect{E}^1(z,t)$ and $\vect{E}^2(z+\zeta,t+\tau)$. $\eta =\omega \tau - k\zeta$ is defined \msgguy{where $\tau$ and $\zeta$ are respectively the time and optical path differences between the waves collected by telescopes 1 and 2}. Then, $\eta$ is equal to zero at the zero optical path difference, which is the origin of the visibility phase. The two waves are injected in Eq.~\ref{eq:mod}, yielding the following expression for the $x$ axis:
\begin{equation}
\begin{split}
\label{eq:start}
\tilde{I}_{\bullet \mathrm{p},x} = \mathrm{Re}\left[ e^{-i\eta} \left< \right.\right.&  \tau_{x_1}\tau_{x_2}E_x^2\cos\theta_1\cos\theta_2 e^{-i(\psi_{{x_1}}-\psi_{{x_2}})} 
+ \tau_{{y_1}}\tau_{{y_2}}E_y^2\sin\theta_1\sin\theta_2 e^{-i(\psi_{{y_1}}-\psi_{{y_2}})} \\
&\!\!\!\!\!\!\left.\left.- \tau_{x_1}\tau_{{y_2}}E_xE_y\cos\theta_1\sin\theta_2 e^{-i(\varphi_x-\varphi_y+\psi_{{x_1}}-\psi_{{y_2}})} 
- \tau_{x_2}\tau_{{y_1}}E_xE_y\cos\theta_2\sin\theta_1 e^{-i(\varphi_y-\varphi_x+\psi_{{y_1}}-\psi_{{x_2}})} \right>\right],
\end{split}
\end{equation}
I then \msgguy{classically} define \msgguy{$\rho$ and $\phi$ to help identify trigonometric relations in the above equation and its derivations}:
\begin{equation}
\label{eq:rho_phi}
 \left\{\begin{array}{@{}l@{}}
\rho\cos\phi= \sqrt{\tau_{x_1}\tau_{x_2}}E_x \\ 
\rho\sin\phi= \sqrt{\tau_{y_1}\tau_{y_2}}E_y
\end{array}\right.
,\end{equation}
and rewrite the previous equation as \msgguy{(see Sect.~\ref{sec:eq14_det} for the details of the derivation)}:
\begin{equation}
\label{eq:basic_x}
\begin{split}
2\tilde{I}_{\bullet \mathrm{p},x} = \mathrm{Re}\left[ e^{-i(\eta+\Delta \mathrm{A}\psi)} \left< \rho^2 \right. \right.&  \left[ \left[ \cos(\theta_1+\theta_2)\cos2\phi +\cos(\theta_1-\theta_2) \right]\cos(\tfrac{1}{2}\Delta\Delta\psi)+i \left[ \cos(\theta_1+\theta_2)+\cos(\theta_1-\theta_2)\cos2\phi \right]\sin(\tfrac{1}{2}\Delta\Delta\psi)\right]\\
&\!\!\!\!\!\!\!\!\!\!\!\!\!\!\!\!\!\!\!\!\!\!\!\!\!\!\!\!\!\!\!\!\!\!\!\!\!\!\!\!\!\!\!\!\!\!\!\!\!\!\!\!\!\!\!\!\!\!\!\!\!\!\!\!\!\!\!\!\!\!\!\!\!\!\!\!\!\!\!\!\!\!\!\!\left.\left.-E_xE_y \left[ \cos\gamma \left[ (\tau_{x_1}\tau_{{y_2}}+\tau_{x_2}\tau_{{y_1}})\sin(\theta_1+\theta_2) -(\tau_{x_1}\tau_{{y_2}}-\tau_{x_2}\tau_{{y_1}})\sin(\theta_1-\theta_2)\right]
+i\sin\gamma \left[ -(\tau_{x_1}\tau_{{y_2}}-\tau_{x_2}\tau_{{y_1}})\sin(\theta_1+\theta_2) +(\tau_{x_1}\tau_{{y_2}}+\tau_{x_2}\tau_{{y_1}})\sin(\theta_1-\theta_2)\right]\right]\right>\right],
\end{split}
\end{equation}
%
The expression for $\tilde{I}_{\bullet \mathrm{p},y}$ can be easily deduced by adding $\frac{3\pi}{2}$ to $\theta_1$ and $\theta_2$:
\begin{equation}
\label{eq:basic_y}
\begin{split}
2\tilde{I}_{\bullet \mathrm{p},y} = \mathrm{Re}\left[ e^{-i(\eta+\Delta \mathrm{A}\psi)} \left< \rho^2 \right. \right.&  \left[ \left[-\cos(\theta_1+\theta_2)\cos2\phi +\cos(\theta_1-\theta_2) \right]\cos(\tfrac{1}{2}\Delta\Delta\psi)+i \left[-\cos(\theta_1+\theta_2)+\cos(\theta_1-\theta_2)\cos2\phi \right]\sin(\tfrac{1}{2}\Delta\Delta\psi)\right]\\
&\!\!\!\!\!\!\!\!\!\!\!\!\!\!\!\!\!\!\!\!\!\!\!\!\!\!\!\!\!\!\!\!\!\!\!\!\!\!\!\!\!\!\!\!\!\!\!\!\!\!\!\!\!\!\!\!\!\!\!\!\!\!\!\!\!\!\!\!\!\!\!\!\!\!\!\!\!\!\!\!\!\!\!\!\left.\left.+E_xE_y \left[ \cos\gamma \left[(\tau_{x_1}\tau_{{y_2}}+\tau_{x_2}\tau_{{y_1}})\sin(\theta_1+\theta_2) + (\tau_{x_1}\tau_{{y_2}}-\tau_{x_2}\tau_{{y_1}})\sin(\theta_1-\theta_2)\right]
-i\sin\gamma \left[(\tau_{x_1}\tau_{{y_2}}-\tau_{x_2}\tau_{{y_1}})\sin(\theta_1+\theta_2) +(\tau_{x_1}\tau_{{y_2}}+\tau_{x_2}\tau_{{y_1}})\sin(\theta_1-\theta_2)\right]\right]\right>\right].
\end{split}
\end{equation}
%
In the total intensity $\tilde{I}_{\bullet \mathrm{p}}=\tilde{I}_{\bullet \mathrm{p},x}+\tilde{I}_{\bullet \mathrm{p},y}$, all $\theta_1+\theta_2$ terms disappear while the $\theta_1-\theta_2$ terms are doubled, yielding:
\begin{equation}
\label{eq:basic_tot}
\begin{split}
\!\!\!\!\!\!\!\!\!\!\!\!\!\!\! \tilde{I}_{\bullet \mathrm{p}}= \mathrm{Re}\left[ e^{-i(\eta+\Delta \mathrm{A}\psi)}  \left<  \rho^2 \cos(\theta_1 \! -\theta_2)\left[ \cos(\tfrac{1}{2}\Delta\Delta\psi)+i\cos2\phi \sin(\tfrac{1}{2}\Delta\Delta\psi)\right] +E_xE_y \sin(\theta_1 \!-\theta_2)\left[(\tau_{x_1}\tau_{{y_2}}\!\!-\!\tau_{x_2}\tau_{{y_1}})\cos\gamma -i(\tau_{x_1}\tau_{{y_2}}\!\!+\!\tau_{x_2}\tau_{{y_1}})\sin\gamma \right]\right>\right].
\end{split}
\end{equation}
\section{Symmetric interferometer}
\label{sec:symmetric}
%
%
Interferometers are designed to be symmetric to minimize polarization effects in instruments. It means that for any characteristic $\chi$ value of the instrument, $\chi_1=\chi_2$. As a consequence, $\theta_1=\theta_2$, which is natural for maximizing the fringe contrast, and $\tau_{x_1}=\tau_{x_2}=\tau_x$, $\tau_{y_1}=\tau_{y_2}=\tau_y$. This also applies to the retardance induced by optical reflections or thin layers, which can be considered the same in all arms of the interferometer. In the following, \msgguy{we} still consider a source of retardance, namely: the birefringence of propagation media like fibers. As a consequence, the $\Delta \Delta \psi$ term is essentially due to the propagation media, while the A$\Delta \psi$ term is due to both optical surfaces or thin layers and to propagation media. The interferometer is polarizing in the general case but is non-polarizing when, in addition, $\tau_x=\tau_y=\tau$. 
\subsection{Basic expressions}  
%
With these hypotheses, Eqs.~\ref{eq:basic_x}, \ref{eq:basic_y}, and \ref{eq:basic_tot} become:
\begin{equation}
\label{eq:I_x_no_rot}
\begin{split}
2\tilde{I}_{\bullet \mathrm{p},x} = \mathrm{Re}\left[ e^{-i(\eta+\Delta \mathrm{A}\psi)}  \left< \rho^2 \left[ \left[1+\cos2\theta\cos2\phi \right]\cos(\tfrac{1}{2}\Delta\Delta\psi) +i \left[ \cos2\theta+\cos2\phi \right]\sin(\tfrac{1}{2}\Delta\Delta\psi)\right] \right.\right.\\
\left.\left.-E_xE_y\sin2\theta \left[(\tau_{x_1}\tau_{{y_2}}+\tau_{x_2}\tau_{{y_1}})\cos\gamma -i(\tau_{x_1}\tau_{{y_2}}-\tau_{x_2}\tau_{{y_1}})\sin\gamma\right]\right> \right]
\end{split}
,\end{equation}
\begin{equation}
\label{eq:I_y_no_rot}
\begin{split}
2\tilde{I}_{\bullet \mathrm{p},y} = \mathrm{Re}\left[ e^{-i(\eta+\Delta \mathrm{A}\psi)}  \left< \rho^2 \left[ \left[1-\cos2\theta\cos2\phi \right]\cos(\tfrac{1}{2}\Delta\Delta\psi) +i \left[-\cos2\theta+\cos2\phi \right]\sin(\tfrac{1}{2}\Delta\Delta\psi)\right]\right.\right.\\
\msgguy{\left.\left.-E_xE_y \sin2\theta \left[-(\tau_{x_1}\tau_{{y_2}}+\tau_{x_2}\tau_{{y_1}})\cos\gamma +i(\tau_{x_1}\tau_{{y_2}}-\tau_{x_2}\tau_{{y_1}})\sin\gamma\right]\right>\right]}
\end{split}
,\end{equation}
%
\begin{equation}
\label{eq:I_tot_no_rot}
\begin{split}
\tilde{I}_{\bullet \mathrm{p}}= \mathrm{Re}\left[ e^{-i(\eta+\Delta \mathrm{A}\psi)}\left< \rho^2  \cos(\tfrac{1}{2}\Delta\Delta\psi)+i\cos2\phi \sin(\tfrac{1}{2}\Delta\Delta\psi)\right>\right].
\end{split}
\end{equation}
Equation~(\ref{eq:rho_phi}) is now written as:
%
%
\begin{equation}
\label{eq:rho_phi_pol}
 \left\{\begin{array}{@{}l@{}}
\rho\cos\phi= \tau_x E_x \\ 
\rho\sin\phi= \tau_y E_y
\end{array}\right.
.\end{equation}
Then $\rho^2 = \tau^2_x E_x^2+\tau^2_yE_y^2$ and $\rho^2\cos2\phi=\tau^2_xE_x^2-\tau^2_yE_y^2$ with the special cases:
\begin{equation}
\label{eq:rho_phi_pol}
 \begin{cases}
 \mathrm{standard\,\,light\!:} & \;\;<\rho^2> = (\tau^2_x+\tau^2_y) I_\mathrm{np} /2 \,, \,\,\,\, \,\,\,\,\,\, \,\,\,\, \,\,\,\,\,\, \,\,\,\, \,\,\,\,\,\, \,\,\,\, \,  <\rho^2 \cos 2\phi>=(\tau^2_x-\tau^2_y) I_\mathrm{np} /2,\\ 
\mathrm{polarized\,\,light\!:} & \,\,\,\,\,\,\,\,\, \rho^2 \;\;=(\tau^2_x\cos^2\alpha+\tau^2_y\sin^2\alpha) I_\mathrm{p}\, , \,\,\,\,\,\,\,\,\,\,  \,\,\,\,\,\,\,   \rho^2\cos 2\phi \, \;\;=(\tau^2_x\cos^2\alpha-\tau^2_y\sin^2\alpha) I_\mathrm{p}. 
\end{cases}
\end{equation}
%
%
%
%
The expressions of the interferograms can now be derived for the two kinds of light from Eqs.~\ref{eq:I_x_no_rot}, \ref{eq:I_y_no_rot}, and \ref{eq:I_tot_no_rot}. For standard light:
\begin{equation}
\label{eq:I_no_rot_sym_non-pol}
 \left\{\begin{array}{@{}l@{}}
4\tilde{I}_{\mathrm{np},x} = I_\mathrm{np}\bigl[\cos(\eta+\Delta \mathrm{A}\psi) [ (\tau_x^2+\tau_y^2)+(\tau_x^2-\tau_y^2)\cos2\theta] \cos(\tfrac{1}{2}\Delta\Delta\psi) + \sin(\eta+\Delta \mathrm{A}\psi)  [(\tau_x^2-\tau_y^2)+(\tau_x^2+\tau_y^2)\cos2\theta]\sin(\tfrac{1}{2}\Delta\Delta\psi)\bigr],  \\
4\tilde{I}_{\mathrm{np},y} = I_\mathrm{np}\bigl[\cos(\eta+\Delta \mathrm{A}\psi) [ (\tau_x^2+\tau_y^2)-(\tau_x^2-\tau_y^2)\cos2\theta] \cos(\tfrac{1}{2}\Delta\Delta\psi) + \sin(\eta+\Delta \mathrm{A}\psi)  [(\tau_x^2-\tau_y^2)-(\tau_x^2+\tau_y^2)\cos2\theta]\sin(\tfrac{1}{2}\Delta\Delta\psi)\bigr],  \\
%
%
2\tilde{I}_\mathrm{np} \,\,\,= I_\mathrm{np}\bigl[\cos(\eta+\Delta \mathrm{A}\psi) (\tau_x^2+\tau_y^2) \cos(\tfrac{1}{2}\Delta\Delta\psi) + \sin(\eta+\Delta \mathrm{A}\psi)  (\tau_x^2-\tau_y^2)\sin(\tfrac{1}{2}\Delta\Delta\psi)\bigr].  \\
%
\end{array}\right.
\end{equation}
In the case of a non-polarizing interferometer, namely, for $\tau_x=\tau_y=\tau$, and for unpolarized light, we would get the traditional result \redguy{that} the effect of retardance is to reduce the fringe contrast (see e.g., \citet{Perraut1996} with $\tau=1$), so that the fringe contrasts goes down to to 0 when $\Delta\Delta\psi$ is half a fringe or $\pi,$ \redguy{meaning} a bright fringe on one polarization axis exactly coincides with a dark fringe on the other one:
\begin{equation}
\label{eq:ex}
I_\mathrm{np} = \tau^2 I_{\mathrm{np},0}+\tau^2 I_{\mathrm{np},0}+2 \tau^2 I_{\mathrm{np},0}\cos(\tfrac{1}{2}\Delta\Delta\psi)\cos(\eta+\Delta \mathrm{A}\psi) 
.\end{equation}
%
%
For polarized light:
\begin{equation}
\label{eq:I_no_rot_sym_lin-pol}
 \left\{\begin{array}{ll}
4\tilde{I}_{x}  = & \!\!\!\! I_\mathrm{p}\bigl[\cos(\eta \!+\! \Delta \mathrm{A}\psi) 
\bigl[ \bigl[ (\tau^2_x\!+\!\tau^2_y)(1\!+\!\cos2\theta\cos2\alpha)+(\tau^2_x\!-\!\tau^2_y)(\cos2\alpha\!+\!\cos2\theta)\bigr]\cos(\tfrac{1}{2}\Delta\Delta\psi) \!-\! 2\tau_x\tau_y\sin2\theta\sin2\alpha\cos\gamma \bigr] \bigr. \\
&\bigl.+\! \sin(\eta \! +\! \Delta \mathrm{A}\psi)  \bigl[ (\tau^2_x\!-\!\tau^2_y)(1\!+\!\cos2\theta\cos2\alpha)+(\tau^2_x\!+\!\tau^2_y)(\cos2\alpha\!+\!\cos2\theta)\bigr] \sin(\tfrac{1}{2}\Delta\Delta\psi)\bigr], \\
4\tilde{I}_{y}  = & \!\!\!\! I_\mathrm{p}\bigl[\cos(\eta \!+\! \Delta \mathrm{A}\psi) 
\bigl[ \bigl[ (\tau^2_x\!+\!\tau^2_y)(1\!-\!\cos2\theta\cos2\alpha)+(\tau^2_x\!-\!\tau^2_y)(\cos2\alpha\!-\!\cos2\theta)\bigr]\cos(\tfrac{1}{2}\Delta\Delta\psi) \!+\! 2\tau_x\tau_y\sin2\theta\sin2\alpha\cos\gamma \bigr] \bigr. \\
&\bigl.+\! \sin(\eta \! +\! \Delta \mathrm{A}\psi)  \bigl[ (\tau^2_x\!-\!\tau^2_y)(1\!-\!\cos2\theta\cos2\alpha)+(\tau^2_x\!+\!\tau^2_y)(\cos2\alpha\!-\!\cos2\theta)\bigr] \sin(\tfrac{1}{2}\Delta\Delta\psi)\bigr], \\
2\tilde{I}  \,= & \!\!\!\!  I_\mathrm{p}\bigl[\cos(\eta \!+\! \Delta \mathrm{A}\psi) 
\bigl[ (\tau^2_x\!+\!\tau^2_y)+(\tau^2_x\!-\!\tau^2_y)\cos2\alpha\bigr]\cos(\tfrac{1}{2}\Delta\Delta\psi) +\! \sin(\eta \! +\! \Delta \mathrm{A}\psi)  \bigl[ (\tau^2_x\!-\!\tau^2_y)+(\tau^2_x\!+\!\tau^2_y)\cos2\alpha \bigr] \sin(\tfrac{1}{2}\Delta\Delta\psi)\bigr]. 
\end{array}\right.
\end{equation}
%
%
with $\alpha = \frac{\pi}{4}$ and \msgguy{$\gamma = \pm \frac{\pi}{2} +\mathrm{A} \Delta \psi$}  for circular polarization and \msgguy{$\gamma = \mathrm{A} \Delta \psi$} for linear polarization. \msgguy{I remark that the phase shift term, $\Delta\mathrm{A}\psi$, is independent of the degree of polarization of the source and is therefore not considered as a source of phase shift in what follows. This is because this phase shift is the same, no matter the characteristics of the source, and is compensated for all sources by a simple delay.} 
Some first conclusions can be drawn from these equations. First of all, all fringe patterns on the $x$ and $y$ axes are shifted by retardance with respect to the position of the central fringe in absence of retardance \msgguy{because of the presence of the $\sin(\eta+\Delta\mathrm{A}\psi)$ term}. The shifts are opposite on the $x$ and $y$ axes for standard light \msgguy{for a non-polarizing interferometer (i.e., with $\tau_x = \tau_y$) as the $\sin(\eta+\Delta\mathrm{A}\psi)$ terms have opposite signs, while the signs of the $\cos(\eta+\Delta\mathrm{A}\psi)$ term are the same}. For both standard and circularly polarized light \msgguy{($\cos2\alpha=0$)}, the fringes of the combined $x$ and $y$ polarization axes are not shifted \msgguy{for the same type of interferometer, as the $\sin(\eta+\Delta\mathrm{A}\psi)$ term vanishes}. In both cases, the only consequence of retardance on the total light interferogram is a decrease in the fringe contrast\msgguy{, as in the example of Eq.~\ref{eq:ex}}. 
\msgguy{For linearly polarized light, the combined light interferogram is phase shifted in the general case because of retardance.} 
%
%
\subsection{Partially polarized interferograms}
Equations~\ref{eq:I_no_rot_sym_non-pol} and \ref{eq:I_no_rot_sym_lin-pol} are combined to derive the expressions of the interferograms in partially polarized light conditions: 
\begin{equation}
\label{eq:I_no_rot_sym_polarizing}
 \left\{\begin{array}{ll}
2\tilde{I}_{x}  = & \!\!\!\! I_{0}\bigl[\cos(\eta \!+\! \Delta \mathrm{A}\psi) 
\bigl[ \tfrac{1}{2}\bigl[ (\tau^2_x\!+\!\tau^2_y)(1\!+\!P\cos2\theta\cos2\alpha)+(\tau^2_x\!-\!\tau^2_y)(P\cos2\alpha\!+\!\cos2\theta)\bigr]\cos(\tfrac{1}{2}\Delta\Delta\psi) \!-\! P\tau_x\tau_y\sin2\theta\sin2\alpha\cos\gamma \bigr] \bigr. \\
&\bigl.+\! \sin(\eta \! +\! \Delta \mathrm{A}\psi)  \tfrac{1}{2}\bigl[ (\tau^2_x\!-\!\tau^2_y)(1\!+\!P\cos2\theta\cos2\alpha)+(\tau^2_x\!+\!\tau^2_y)(P\cos2\alpha\!+\!\cos2\theta)\bigr] \sin(\tfrac{1}{2}\Delta\Delta\psi)\bigr], \\
2\tilde{I}_{y}  = & \!\!\!\! I_{0}\bigl[\cos(\eta \!+\! \Delta \mathrm{A}\psi) 
\bigl[ \tfrac{1}{2}\bigl[ (\tau^2_x\!+\!\tau^2_y)(1\!-\!P\cos2\theta\cos2\alpha)+(\tau^2_x\!-\!\tau^2_y)(P\cos2\alpha\!-\!\cos2\theta)\bigr]\cos(\tfrac{1}{2}\Delta\Delta\psi) \!+\! P\tau_x\tau_y\sin2\theta\sin2\alpha\cos\gamma \bigr] \bigr. \\
&\bigl.+\! \sin(\eta \! +\! \Delta \mathrm{A}\psi)  \tfrac{1}{2}\bigl[ (\tau^2_x\!-\!\tau^2_y)(1\!-\!P\cos2\theta\cos2\alpha)+(\tau^2_x\!+\!\tau^2_y)(P\cos2\alpha\!-\!\cos2\theta)\bigr] \sin(\tfrac{1}{2}\Delta\Delta\psi)\bigr], \\
\,\,\,\,\tilde{I}  \,= & \!\!\!\!  I_{0}\bigl[\cos(\eta \!+\! \Delta \mathrm{A}\psi) 
\tfrac{1}{2}\bigl[ (\tau^2_x\!+\!\tau^2_y)+(\tau^2_x\!-\!\tau^2_y)P\cos2\alpha\bigr]\cos(\tfrac{1}{2}\Delta\Delta\psi) +\! \sin(\eta \! +\! \Delta \mathrm{A}\psi)  \tfrac{1}{2}\bigl[ (\tau^2_x\!-\!\tau^2_y)+(\tau^2_x\!+\!\tau^2_y)P\cos2\alpha \bigr] \sin(\tfrac{1}{2}\Delta\Delta\psi)\bigr]. 
\end{array}\right.
\end{equation}
All interferograms can be written as:
\begin{equation}
I=\tau^2I_{0}\rho \cos(\eta + \Delta \mathrm{A}\psi-\xi)
\label{eq:interf_point_source}
,\end{equation}
where $\xi$  is the phase shift due to retardance \msgguy{(both $\rho$ and $\xi$ depend on the polarization characteristics of the instrument and on the degree of polarization of the source)}, yielding: 
%
\begin{equation}
\label{eq:xi_polarizing}
 \left\{\begin{array}{@{}l@{}}
2^{|\varepsilon|}\rho\cos \xi= \tfrac{1}{2}\bigl[ (\tau^2_x\!+\!\tau^2_y)(1\!+\!\varepsilon P\cos2\theta\cos2\alpha)+(\tau^2_x\!-\!\tau^2_y)(P\cos2\alpha\!+\!\varepsilon\cos2\theta)\bigr]\cos(\tfrac{1}{2}\Delta\Delta\psi) \!-\! \varepsilon P\tau_x\tau_y\sin2\theta\sin2\alpha\cos\gamma \\
2^{|\varepsilon|}\rho \sin \xi =\tfrac{1}{2}\bigl[ (\tau^2_x\!-\!\tau^2_y)(1\!+\!\varepsilon P\cos2\theta\cos2\alpha)+(\tau^2_x\!+\!\tau^2_y)(P\cos2\alpha\!+\!\varepsilon\cos2\theta)\bigr] \sin(\tfrac{1}{2}\Delta\Delta\psi)
\end{array}\right.
\end{equation}
with $\varepsilon = +1$ for the $x$ axis, $\varepsilon = -1$ for the $y$ axis and $\varepsilon = 0$ for combined polarizations.

Some conclusions can be drawn from these expressions. 
The general conclusion is that differential retardance induces a shift, even in the case of natural light\msgguy{, since $\sin\xi$ is not zero.} In addition, the  amount \redguy{of differential retardance} depends on various parameters: the rotation of the beams, the degree and nature of polarization, and the orientation of linear polarization. 
All cases are summarized in Table~\ref{ref1} 
with some other particular cases.
The phase shift is always proportional to retardance if 
retardance 
\msgguy{is}
 small \msgguy{since $\sin(\tfrac{1}{2}\Delta\Delta\psi)\simeq\tfrac{1}{2}\Delta\Delta\psi$ to the first order in this case}. The rotation of the beams plays a particular role: if the beams  are aligned with the polarization axes ($\theta=0$ or $\pm\frac{\pi}{2}$), then the tangent of the phase shift is always proportional to $\tan(\tfrac{1}{2}\Delta\Delta\psi)$. 
If the beams are rotated by $\frac{\pi}{4}$ with respect to the $(x,y)$ frame, then there is no phase shift for a non-polarizing interferometer \msgguy{(or equivalently with 0 diattenuation, i.e., with $\tau_x = \tau_y$)} \msgguy{and} with natural or partially circularly polarized light \msgguy{($\cos2\alpha=0$)} \msgguy{since $\sin\xi=0$ in this case}. The phase shift is never 0 if diattenuation is present and shifts on the $x$ and $y$ axes are never opposite in the general case. However, the tangent of the phase shift is proportional to $\tan(\tfrac{1}{2}\Delta\Delta\psi)$ for a non-polarized source 
or in combined mode; this means that the phase shift is proportional to retardance to the first order in these cases. 
\msgguy{Finally}, the phase shifts are opposite on the $x$ and $y$ axes for natural light 
for a non-polarizing interferometer, with a zero phase shift in combined mode. 

\vspace{0.5cm}
\noindent\msgguy{{\bf Remark:} In the case of extended sources, the interferograms from individual point-like sources of Eq.~\ref{eq:interf_point_source} need to be integrated over the sky coordinates $\alpha$ and $\delta$ with a component of $\eta$ being the phase term $-2\pi(\alpha u +\delta v)$, where $u$ and $v$ are the spatial frequency coordinates conjugated to the sky coordinates. As a consequence, the visibilities need to be calculated with a modified version of the Zernike-van Cittert theorem where the phase shift due to retardance $\xi$ is added to the classical phase term $2\pi(\alpha u +\delta v)$. In the particular case of sources whose degree of polarization is independent of location, the visibility phase is simply shifted by $\xi$ and a calibration is required. }

\begin{table*}[]
\label{ref1}
\caption{Visibility phase shifts.}

\hspace{-5mm}\begin{tabular}{c|lccc}
\hline \hline
Diatt. & Partial & $x$ & $y$ & Total \\
& polarization & $(\varepsilon = +1)$ & $(\varepsilon = -1)$ & ($\varepsilon = 0$)  \\
\hline
\multirow{3}{*}{$\tau_x=\tau_y$} 
                                                & $P=0$  
& \scalebox{0.75}{$\mathrm{atan} \left(\cos2\theta\tan(\tfrac{1}{2}\Delta\Delta\psi)\right)$} & \scalebox{0.75}{$-\mathrm{atan} \left(\cos2\theta\tan(\tfrac{1}{2}\Delta\Delta\psi)\right)$} & \scalebox{0.75}{0}   \\
                                                 & circular 
& \scalebox{0.75}{$\mathrm{atan}\left(\frac{\cos2\theta\sin(\tfrac{1}{2}\Delta\Delta\psi)}{\cos(\tfrac{1}{2}\Delta\Delta\psi) - P\sin2\theta\cos\gamma}\right)$} & \scalebox{0.75}{$-\mathrm{atan}\left(\frac{\cos2\theta\sin(\tfrac{1}{2}\Delta\Delta\psi)}{\cos(\tfrac{1}{2}\Delta\Delta\psi) + P\sin2\theta\cos\gamma}\right)$} & \scalebox{0.75}{0}  \\
                                                 & linear    
& \scalebox{0.5}{$\mathrm{atan}\left(\frac{\displaystyle [P\cos2\alpha+\cos2\theta]\sin(\tfrac{1}{2}\Delta\Delta\psi)}{\displaystyle [1+P\cos2\theta\cos2\alpha]\cos(\tfrac{1}{2}\Delta\Delta\psi) - P\sin2\theta\sin2\alpha\cos\gamma}\right)$} 
& \scalebox{0.5}{$\mathrm{atan}\left(\frac{\displaystyle [P\cos2\alpha-\cos2\theta]\sin(\tfrac{1}{2}\Delta\Delta\psi)}{\displaystyle [1-P\cos2\theta\cos2\alpha]\cos(\tfrac{1}{2}\Delta\Delta\psi)+P\sin2\theta\sin2\alpha\cos\gamma}\right)$}  
& \scalebox{0.75}{$\mathrm{atan} \left(P\cos2\alpha\tan(\tfrac{1}{2}\Delta\Delta\psi)\right)$} \\
\hline
\multirow{3}{*}{$\tau_x\neq\tau_y$}  
%
%
%
                                                & $P=0$   
&\!\!\!\! \!\!\!\! \!\!\!\! \!\!\!\! \!\!\!\! \!\!\!\! \!\!\!\! \scalebox{0.7}{$\mathrm{atan}\left(\frac{\displaystyle [\tau_x^2\cos^2\theta-\tau_y^2\sin^2\theta]}{\displaystyle [\tau_x^2\cos^2\theta+\tau_y^2\sin^2\theta]} \tan(\tfrac{1}{2}\Delta\Delta\psi) \right)$}
&\!\!\!\! \!\!\!\! \scalebox{0.7}{$\mathrm{atan}\left(\frac{\displaystyle [\tau_x^2\sin^2\theta -\tau_y^2\cos^2\theta]}{\displaystyle [\tau_x^2\sin^2\theta +\tau_y^2\cos^2\theta]} \tan(\tfrac{1}{2}\Delta\Delta\psi) \right)$}  
& \scalebox{0.7}{$\mathrm{atan} \left(\frac{\displaystyle(\tau_x^2-\tau_y^2)}{\displaystyle (\tau_x^2+\tau_y^2)}\tan(\tfrac{1}{2}\Delta\Delta\psi)\right)$} \\
                                                 & circ.      
&\!\!\!\! \!\!\!\! \!\!\!\! \!\!\!\! \!\!\!\! \!\!\!\! \!\!\!\! \scalebox{0.7}{$\mathrm{atan}\left(\frac{\displaystyle [\tau_x^2\cos^2\theta -\tau_y^2\sin^2\theta]\sin(\tfrac{1}{2}\Delta\Delta\psi)}{\displaystyle [\tau_x^2\cos^2\theta+\tau_y^2\sin^2\theta]\cos(\tfrac{1}{2}\Delta\Delta\psi) - P\sin2\theta\cos\gamma}\right)$} 
&\!\!\!\! \!\!\!\! \scalebox{0.7}{$\mathrm{atan}\left(\frac{\displaystyle [\tau_x^2\sin^2\theta-\tau_y^2\cos^2\theta]\sin(\tfrac{1}{2}\Delta\Delta\psi)}{\displaystyle [\tau_x^2\sin^2\theta+\tau_y^2\cos^2\theta]\cos(\tfrac{1}{2}\Delta\Delta\psi) + P\sin2\theta\cos\gamma}\right)$}  
& \scalebox{0.7}{$\mathrm{atan} \left(\frac{\displaystyle(\tau_x^2-\tau_y^2)}{\displaystyle (\tau_x^2+\tau_y^2)}\tan(\tfrac{1}{2}\Delta\Delta\psi)\right)$} \\
                                                 & lin.   
&\!\!\!\! \!\!\!\! \!\!\!\! \!\!\!\! \!\!\!\! \!\!\!\! \!\!\!\! \scalebox{0.5}{$\mathrm{atan}\left(\frac{\displaystyle [\tau_x^2\cos^2\theta(1+P\cos2\alpha)-\tau_y^2\sin^2\theta(1-P\cos2\alpha)]\sin(\tfrac{1}{2}\Delta\Delta\psi)}{\displaystyle [\tau_x^2\cos^2\theta(1+P\cos2\alpha)+\tau_y^2\sin^2\theta(1-P\cos2\alpha)]\cos(\tfrac{1}{2}\Delta\Delta\psi) - P\sin2\theta\sin2\alpha\cos\gamma}\right)$} 
&\!\!\!\! \!\!\!\! \scalebox{0.5}{$\mathrm{atan}\left(\frac{\displaystyle [\tau_x^2\sin^2\theta(1+P\cos2\alpha)-\tau_y^2\cos^2\theta(1-P\cos2\alpha)]\sin(\tfrac{1}{2}\Delta\Delta\psi)}{\displaystyle [\tau_x^2\sin^2\theta(1+P\cos2\alpha)+\tau_y^2\cos^2\theta(1-P\cos2\alpha)]\cos(\tfrac{1}{2}\Delta\Delta\psi) + P\sin2\theta\sin2\alpha\cos\gamma}\right)$}  
& \scalebox{0.58}{$\mathrm{atan} \left(\frac{\displaystyle(\tau_x^2-\tau_y^2)+(\tau_x^2+\tau_y^2)P\cos2\alpha}{\displaystyle (\tau_x^2+\tau_y^2)+(\tau_x^2-\tau_y^2)P\cos2\alpha}\tan(\tfrac{1}{2}\Delta\Delta\psi)\right)$} \\
\hline
\end{tabular}
\tablefoot{Visibility phase shift as a function of the diattenuation properties of the interferometer, of the polarization of the source and of the output polarization axis (either split or combined polarizations). The expressions are the direct application of Eq.~(\ref{eq:xi_polarizing}).}
\end{table*}
%
%
\section{Biases of the visibility phase and derived quantities}
\label{sec:phase_bias}
I  have shown (see Sect.~\ref{sec:symmetric}) that in the general case, the phase of the visibility is biased in presence of retardance whatever the polarization characteristics of the source; however, in some cases, the phase bias is strictly zero (as shown in  Table~\ref{ref1}). 
Here, the biases on differential and closure phases are discussed in the general case.
\paragraph{Differential phase}If the chromaticity of the phase shift \msgguy{$\xi$ in Eq.~\ref{eq:xi_polarizing}} is negligible, then the differential phase can be considered to be unbiased by the polarization properties of the interferometer.
\paragraph{Closure phase}The closure phase can be immune to retardance as well under some circumstances. We can introduce, $\tau,  $ which is the average transmission on the two polarization axes and $\delta\tau=\tau_x-\tau_y$ is the difference in transmission between the two axes.  For weak diattenuation and differential retardance, as well as for moderately polarized sources (i.e., assuming the degree of polarization is a few tens of percent at most, which is quite standard for polarized astronomical sources), the Eq.~\ref{eq:xi_polarizing} system can be expanded to the second order in $P$, $\Delta \Delta \psi$ and $\delta\tau$ yielding for each baseline $i$:
\begin{equation}
\label{eq:phase_shift}
\xi_i \approx \frac{1}{4}\left[2\varepsilon\cos 2\theta +2(P\cos 2\alpha+\frac{\delta\tau}{\tau})(1-\varepsilon^2\cos^2 2\theta)
+\varepsilon^2P\sin 4\theta\sin 2\alpha\cos\gamma_i\right] \Delta \Delta \psi_i
.\end{equation}
If the differential retardances $\Delta\Delta\psi_i$ are first order quantities then the average retardances $\mathrm{A}\Delta\psi_i$ are a same constant plus first order quantities linearly linked to the $\Delta\Delta\psi_i$ so that 
$\mathrm{A}\Delta\psi_i = \gamma_0 +\left( \frac{1}{2}-\frac{2}{N}\right)\sum_{j=1...N}\left(\Delta\psi_{T^2_i} - \Delta\psi_j +  \Delta\psi_{T^1_i} -\Delta\psi_j  \right), $ where $T^1_i$ and $T^2_i$ are, respectively, the first and second telescopes of baseline $i$,
%
$\gamma_0=\frac{1}{N}\sum_{j=1...N}{\Delta\psi_j}$\msgguy{, $N$ being the number of baselines}. As a consequence, $\gamma_i \approx \gamma = (\varphi_x-\varphi_y) + \gamma_0$ to zero order. That means that the phase bias \redguy{writes} to the second order:
\begin{equation}
\label{eq:xi_expansion}
\xi_i \approx \frac{1}{4}\left[2\varepsilon\cos 2\theta +2(P\cos 2\alpha+\frac{\delta\tau}{\tau})(1-\varepsilon^2\cos^2 2\theta)
+\varepsilon^2P\sin 4\theta\sin 2\alpha\cos\gamma\right] \Delta \Delta \psi_i
.\end{equation}
Since the $\Delta \Delta \psi_i$ are closing quantities and the term between brackets is independent of the baseline, the $\xi_i$ values are also closing quantities to the second order. Here, the consequence is that the bias on closure phases is zero to this order. 
%

%
%
%
%
%
%
\section{Astrometric bias}
\label{sec:astrometry}
%
%
\subsection{Theoretical derivation}
We considered a phase-referenced interferometer for which the phase of the visibility of a point source is equal to 0 at the center of the coordinate frame. In such a case, the visibility of a point source located at $\vect{\alpha}$ writes $V(\vect{u})=e^{-2i\pi(\alpha u + \delta v)}$. The position of the point source is deduced from the visibility phases measured by the interferometer. One way to deduce one from the other is to find the optimum phase model $\varphi_{\vect{\alpha}}(\vect{u})=2\pi(\alpha u + \delta v)$ that minimizes the quantity:
\begin{equation}
\label{eq:chi2}
\chi^2(\vect{\alpha})=\sum_{i=1}^{N}\left[ \xi_i-2\pi(\alpha u_i + \delta v_i) \right]^2
,\end{equation}
where $N$ is the number of baselines or the number of visibility points available if they can be  
combined over time. Since the phase model is linear, there is a linear relation between the interferogram phases and the optimum position:
\begin{equation}
\label{eq:opt_sol}
                \left[\begin{array}{l}
                       \alpha   \\
                       \delta
                     \end{array} \right] =M^{-1}\left[\begin{array}{l}
                       \sum_{i=1}^{N} \xi_i u_i    \\
                       \sum_{i=1}^{N} \xi_i v_i 
                     \end{array} \right] 
,\end{equation}
where:
\begin{equation}
                M=2\pi\left[\begin{array}{ll}
                       \sum_{i=1}^{N} u_i^2 & \sum_{i=1}^{N} u_i v_i    \\
                      \sum_{i=1}^{N} u_i v_i & \sum_{i=1}^{N} v_i^2
                     \end{array} \right] 
,\end{equation}
and:
\begin{equation}
                M^{-1}=\frac{2\pi}{\mathrm{det}\,M}\left[\begin{array}{ll}
                       \,\,\,\,\, \sum_{i=1}^{N} v_i^2 & -\sum_{i=1}^{N} u_i v_i    \\
                      -\sum_{i=1}^{N} u_i v_i & \,\,\,\,\, \sum_{i=1}^{N} u_i^2
                     \end{array} \right] 
,\end{equation}
with $\mathrm{det}\,M = 4\pi^2\left[ \left( \sum_{i=1}^{N} u_i^2 \right) \left( \sum_{i=1}^{N} v_i^2 \right) -\right( \sum_{i=1}^{N} u_i v_i  \left)^2\right]$. When $\mathrm{det}\,M \neq 0$, the general solution of the linear system is:
%
%
\begin{equation}
\label{eq:sol_bias}
\left[\begin{array}{l}
        \alpha  \\
        \delta
\end{array} \right] = \frac{2\pi}{\mathrm{det}\,M} \left[\begin{array}{l}
        \left( \sum_{i=1}^{N} v_i^2 \right)\left( \sum_{i=1}^{N} \xi_i u_i  \right) -\left( \sum_{i=1}^{N} u_i v_i  \right)\left( \sum_{i=1}^{N} \xi_i v_i  \right)    \\
        \left( \sum_{i=1}^{N} u_i^2 \right)\left( \sum_{i=1}^{N} \xi_i v_i  \right) -\left( \sum_{i=1}^{N} u_i v_i  \right)\left( \sum_{i=1}^{N} \xi_i u_i  \right)
 \end{array} \right] 
.\end{equation}
Let us note $\vect{U}=\{u_i\}_{i=1\dots N}$ and $\vect{V}=\{v_i\}_{i=1\dots N}$, then $\mathrm{det}\,M = 4\pi^2 [ ||\vect{U}||^2||\vect{V}||^2 - ( \vect{U}.\vect{V})^2 ]$. According to the Cauchy-Schwartz theorem, the determinant is always positive and is zero if and only if $\vect{U}$ and $\vect{V}$ are colinear meaning there exists a constant $\kappa$ such that $\vect{U} = \kappa \vect{V}$
. \msgguy{In the general case, }we can write $\vect{U}$ as the sum of its projection on $\vect{V}$ and of a vector perpendicular to $\vect{V}$. This yields: $\vect{U} = \kappa \vect{V} + \vect{\omega}$, where $\kappa=\vect{U}.\vect{V}/||\vect{V}||^2$. The determinant is therefore expressed as: $\mathrm{det}\,M = 4\pi^2 ||\vect{\omega}||^2||\vect{V}||^2$
and coupled with Eq.~\ref{eq:sol_bias}, we have:
\begin{equation}
\label{eq:sol_bias_omega}
\left[\begin{array}{l}
        \alpha  \\
        \delta
\end{array} \right] = \frac{2\pi}{\mathrm{det}\,M} \left[\begin{array}{l}
\;\;\; \;   ||\vect{V}||^2 (\vect{\omega}.\vect{\Xi}) \\
        -\kappa ||\vect{V}||^2 (\vect{\omega}.\vect{\Xi})
        + ||\vect{\omega}||^2 (\vect{V}.\vect{\Xi})
 \end{array} \right] 
 =
  \frac{\vect{\omega}.\vect{\Xi}}{2\pi||\vect{\omega}||^2} \left[\begin{array}{l}
  \,\,  \,\, 1 \\
        -\kappa 
 \end{array} \right] 
 +
  \frac{\vect{V}.\vect{\Xi}}{2\pi||\vect{V}||^2} \left[\begin{array}{l}
0 \\
1
 \end{array} \right] 
\end{equation}
with $\vect{\Xi}=\{\xi_i\}_{i=1\dots N}$. Equivalent expressions are obtained if $\vect{V}$ is projected onto $\vect{U}$ swapping $\alpha \leftrightarrow \delta$ and $\vect{U} \leftrightarrow \vect{V}$. In the case where $\mathrm{det}\,M$ is close to 0, $\vect{\omega}$ is a negligible quantity. The terms of the third side of the equation are ordered so that the second half is of zero order in $||\vect{\omega}||$ and the first half of the vector is of the order of $1/||\vect{\omega}||$ and dominates  except in the particular case where $\vect{\omega}$ is perpendicular to $\vect{\Xi}$ for which only the last term remains. 
Except in this particular case, for a small $\vect{\omega}$, the astrometric bias is approximately colinear to $[1, -\kappa]^T$ and can take arbitrarily large values (it becomes infinite for $\vect{\omega}=\vect{0}$ meaning the astrometry cannot be constrained in this case). 

For $\mathrm{det}\,M \neq 0$, for weak diattenuation and differential retardance and for moderately polarized sources, the expansion of $\xi_i$ to the second order of Eq.~\ref{eq:xi_expansion} can be used 
and injected into Eq.~\ref{eq:opt_sol} to get the astrometric bias expansion to the second order:
%
%
\begin{equation}
\label{eq:sol_bias_lin_bis}
\left[\begin{array}{l}
        \alpha  \\
        \delta
\end{array} \right]
\approx \frac{1}{4}\left[2\varepsilon\cos 2\theta +2(P\cos 2\alpha+\frac{\delta\tau}{\tau})(1-\varepsilon^2\cos^2 2\theta)
+\varepsilon^2P\sin 4\theta\sin 2\alpha\cos\gamma\right] 
\left[\begin{array}{l}
        \alpha_0    \\
         \delta_0
 \end{array} \right], 
\end{equation}
where:
\begin{equation}
\label{eq:alphas}
\vect{\alpha_0} = M^{-1}\left[\begin{array}{l}
        \sum_{i=1}^{N} \Delta\Delta\psi_i u_i    \\
         \sum_{i=1}^{N} \Delta\Delta\psi_i v_i  
 \end{array} \right].\\
\end{equation}
\begin{table*}[t]
\label{tab:astrometric_bias}
\caption{Astrometric bias.}
\hspace{-5mm}\begin{tabular}{c|lccc}
\hline \hline
Diatt. & Partial & $x$ & $y$ & Total \\
& polarization & $(\varepsilon = +1)$ & $(\varepsilon = -1)$ & ($\varepsilon = 0$)  \\
\hline
\multirow{3}{*}{$\delta\tau=0$} & $P=0$  & 
\scalebox{0.75}{$\frac{1}{2}\cos 2\theta
\,\vect{\alpha_0}$}%
 &
\scalebox{0.75}{$-\frac{1}{2}\cos 2\theta
\,\vect{\alpha_0}$}%
 & \scalebox{0.75}{0}   \\
                                                 & circular & 
\scalebox{0.75}{$\frac{1}{4}\left[2\cos 2\theta + P\sin 4\theta\cos\gamma\right] 
\vect{\alpha_0}$}%
                 & 
\scalebox{0.75}{$\frac{1}{4}\left[-2\cos 2\theta + P\sin 4\theta\cos\gamma\right] 
\vect{\alpha_0}$}%
& \scalebox{0.75}{0}  \\
                                                 & linear    & 
\!\!\!\! \!\!\!\! \!\!\!\! \!\!\!\! \!\!\!\!  \scalebox{0.75}{$\frac{1}{4}[2\cos 2\theta +2P\cos 2\alpha\sin^2 2\theta
+P\sin 4\theta\sin 2\alpha\cos\gamma]
\vect{\alpha_0}$}%
          &
\scalebox{0.75}{$\frac{1}{4}\left[-2\cos 2\theta +2P\cos 2\alpha\sin^2 2\theta
+P\sin 4\theta\sin 2\alpha\cos\gamma\right] 
\vect{\alpha_0}$}%
& 
\scalebox{0.75}{$\frac{1}{2}P\cos 2\alpha
\,\vect{\alpha_0}$}%
  \\
\hline

\multirow{3}{*}{$\delta\tau \neq 0$} & $P=0$  & 
\scalebox{0.75}{$\frac{1}{2}\left[\cos 2\theta +\frac{\delta\tau}{\tau}\sin^2 2\theta \right]
\vect{\alpha_0}$}%
 &
\scalebox{0.75}{$-\frac{1}{2}\left[-\cos 2\theta +\frac{\delta\tau}{\tau}\sin^2 2\theta \right]
\vect{\alpha_0}$}%
 & 
\scalebox{0.75}{$\frac{1}{2}\frac{\delta\tau}{\tau}
\,\vect{\alpha_0}$}%
    \\
                                                 & circular & 
\scalebox{0.75}{$\frac{1}{4}\left[2\cos 2\theta +2\frac{\delta\tau}{\tau}\sin^2 2\theta + P\sin 4\theta\cos\gamma\right] 
\vect{\alpha_0}$}%
                 & 
 \scalebox{0.75}{$\frac{1}{4}\left[-2\cos 2\theta +2\frac{\delta\tau}{\tau}\sin^2 2\theta + P\sin 4\theta\cos\gamma\right] 
\vect{\alpha_0}$}%
&
\scalebox{0.75}{$\frac{1}{2} \frac{\delta\tau}{\tau} 
\vect{\alpha_0}$}%
\\
                                                 & linear    & 
\!\!\!\! \!\!\!\! \!\!\!\! \!\!\!\! \!\!\!\! \!\!\!\! \!\!\!\! \!\!\!\! \scalebox{0.75}{$\frac{1}{4}\left[2\cos 2\theta +2(P\cos 2\alpha+\frac{\delta\tau}{\tau})\sin^2 2\theta
+P\sin 4\theta\sin 2\alpha\cos\gamma\right]
\vect{\alpha_0}$}%
          &
\scalebox{0.75}{$\frac{1}{4}\left[-2\cos 2\theta +2(P\cos 2\alpha+\frac{\delta\tau}{\tau})\sin^2 2\theta
+P\sin 4\theta\sin 2\alpha\cos\gamma\right] 
\vect{\alpha_0}$}%
& 
\scalebox{0.75}{$\frac{1}{2}(P\cos 2\alpha+\frac{\delta\tau}{\tau})
\,\vect{\alpha_0}$}\\
\hline
\end{tabular}
\tablefoot{Astrometric bias as a function of the diattenuation properties of the interferometer, of the polarization of the source and of the output polarization axis (either split or combined polarizations). The expressions are the direct application of Eq~\ref{eq:sol_bias_lin_bis}. Differential retardance, degree of polarization and diattenuation, and supposed to be small.}
\end{table*}
The only terms that depend on the characteristics of the source in Eq.~\ref{eq:sol_bias_lin_bis} are the scalar terms. The vector $\vect{\alpha_0}$  in the second member of the equation only depends on the points in the $(u,v)$ plane and on the retardance (differential and average) of the arms of the interferometer. 
The astrometric bias is therefore proportional to $\vect{\alpha_0}$ to the first order with coefficients depending on the polarization properties of the source, on the $(u,v)$ sampling, and on the polarizing properties of the interferometer. All cases are summarized in Table~\ref{tab:astrometric_bias}.
%
%
%
For a non-polarizing interferometer ($\delta\tau = 0$), there is no astrometric bias for partially circularly polarized light ($\cos 2 \alpha = 0$) when the two polarizations are combined ($\varepsilon = 0$). For linearly polarized light, the astrometric bias is proportional to the degree of polarization $P$ and to $\cos 2 \alpha$ for combined polarizations. If the orientation of the EVPA $\alpha$ varies, then the bias describes a line segment whose inclination is given by $\vect{\alpha_0}$. The astrometries of both split polarizations ($\varepsilon = -1$ and $+1$) are biased for both types of polarizations. The biases vary linearly with the degree $P$ of polarization and are shifted symmetrically by $\pm \frac{1}{2}\, \varepsilon \cos 2\theta \,\vect{\alpha_0}$ on the $x$ and $y$ axes. For a linear polarization, the biases also vary linearly with $P$ and if the EVPA varies by at least an amplitude of $\pi$ or $-\pi$, then the $x$ and $y$ biases cover opposite ranges of values. In this case, the biases also describe  line segments whose inclination is given by $\vect{\alpha_0}$.
%
%
If the interferometer is slightly polarizing ($\delta\tau \neq 0$), the general shape of the biases do not change but an extra source of bias appears because of the $\frac{\delta\tau}{\tau}$ term.
In all cases, the biases can be calibrated knowing the amount of retardance (differential and average) in the arms of the interferometer and the $(u,v)$ plane samples.
\begin{figure}[h]
\includegraphics[width=8cm]{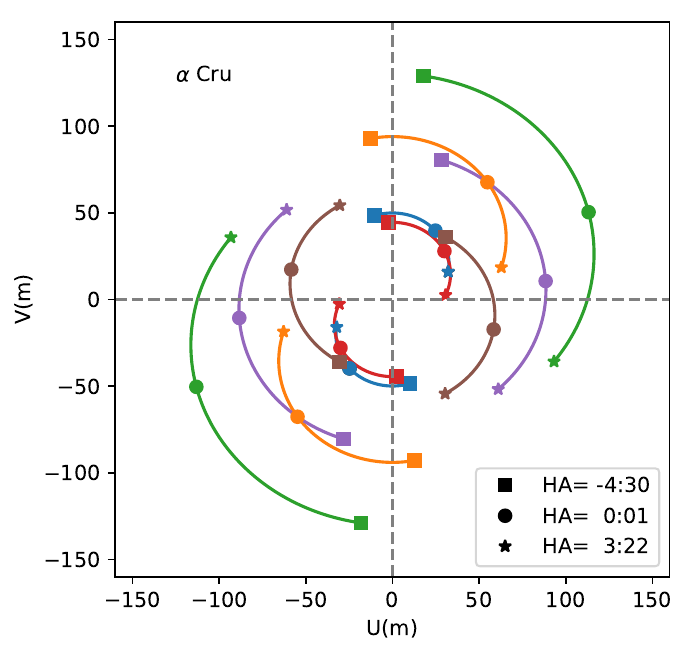} 
\caption{$(u,v)$ plane coverage for $\alpha$ Cru with symbols for samples used for Fig.~\ref{Fig:bias_alpha_Cru}.}
\label{Fig:(u,v)}
\end{figure}
\begin{figure*}[h]
\begin{tabular}{ccc}
\hspace{-0.5cm}\includegraphics[width=6cm]{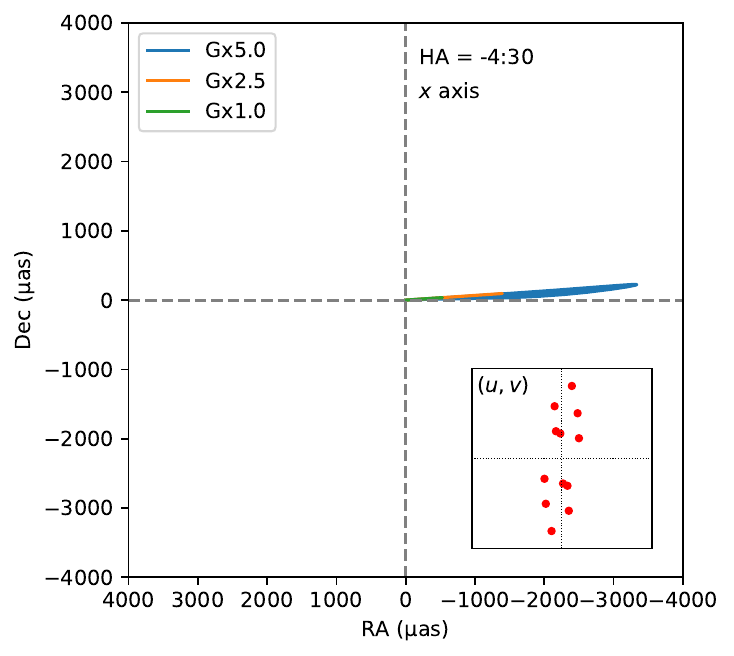} &\includegraphics[width=6cm]{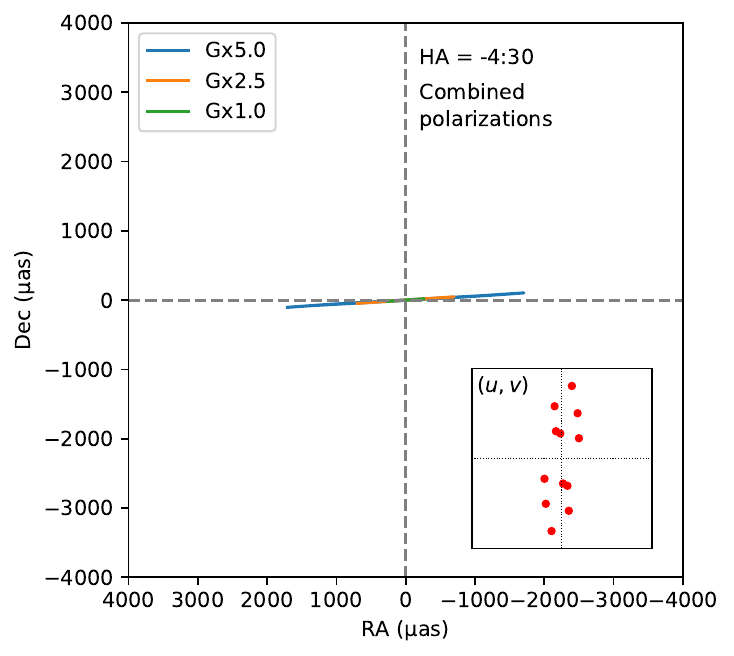} & \includegraphics[width=6cm]{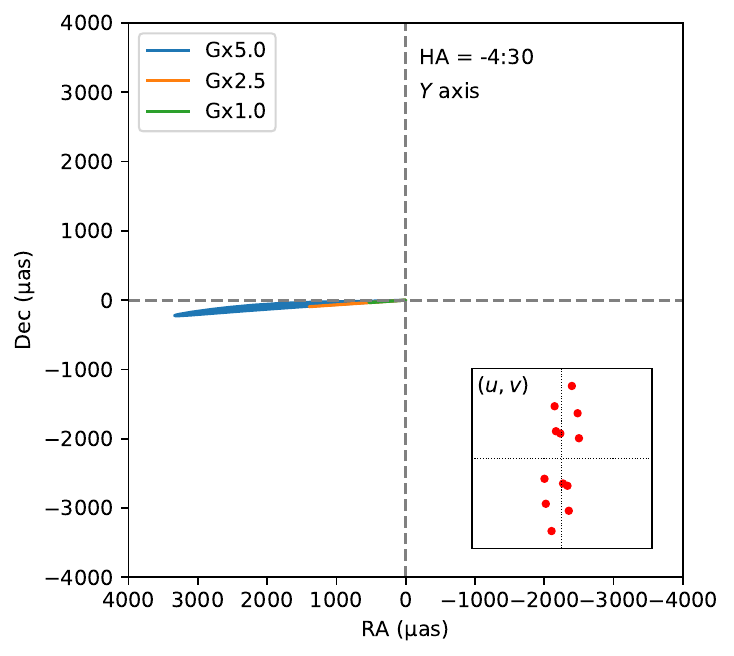} \\
\hspace{-0.5cm}\includegraphics[width=6cm]{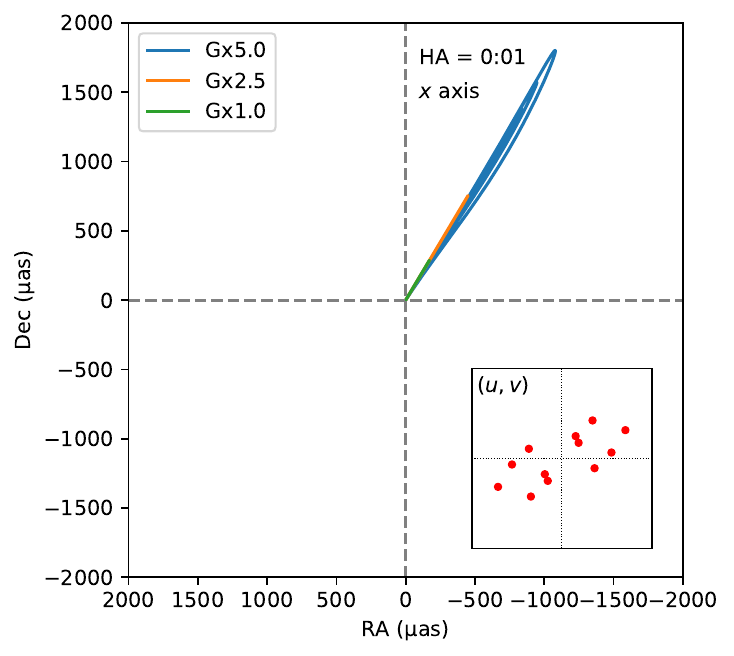} &\includegraphics[width=6cm]{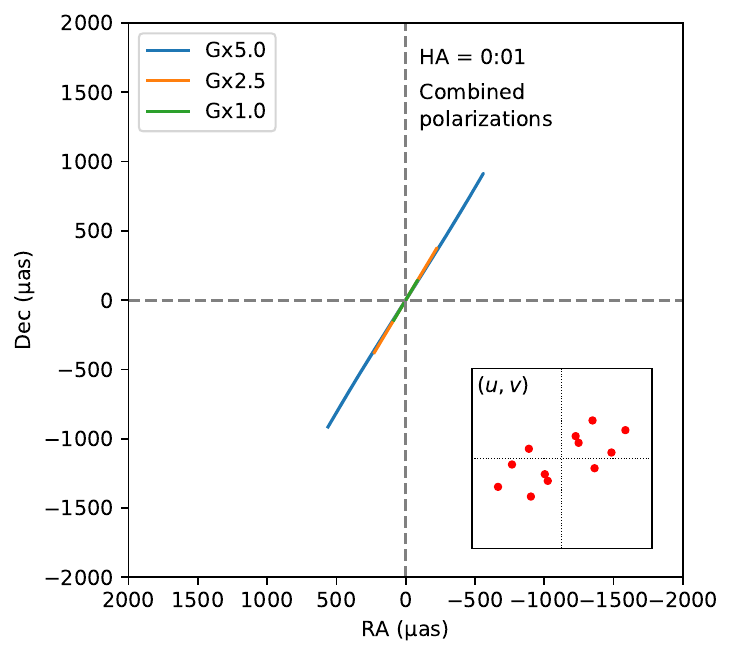} & \includegraphics[width=6cm]{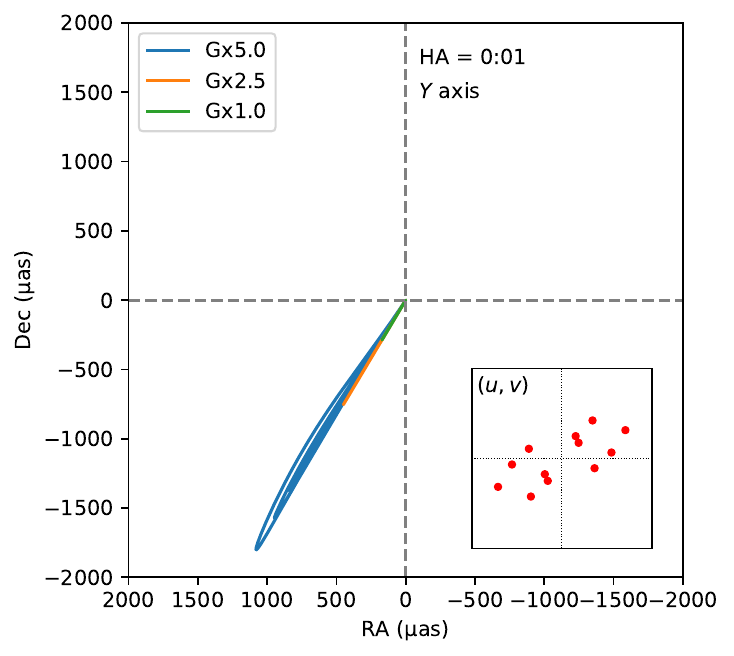} \\
\hspace{-0.5cm}\includegraphics[width=6cm]{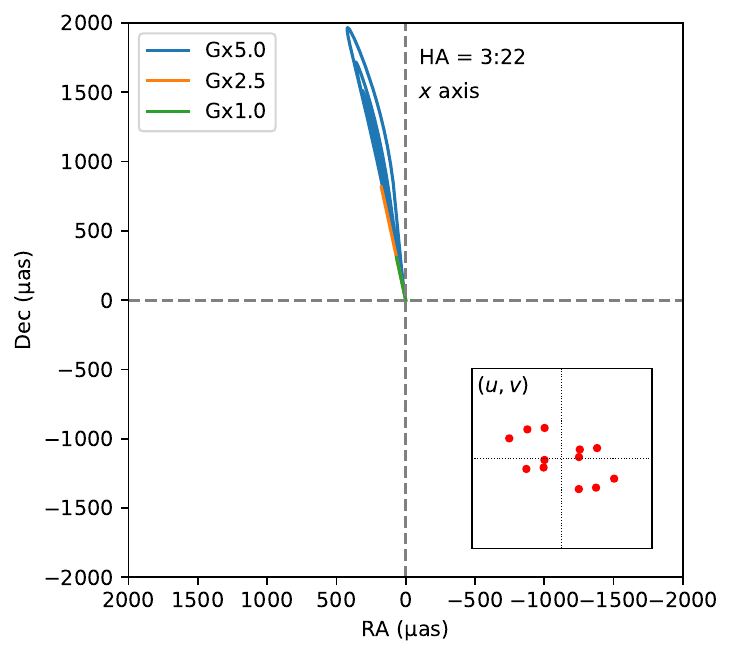} &\includegraphics[width=6cm]{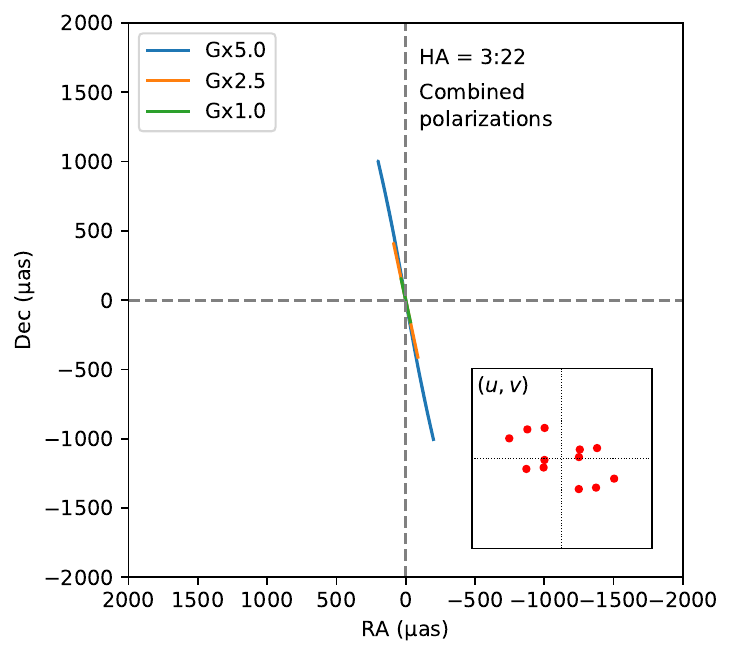} & \includegraphics[width=6cm]{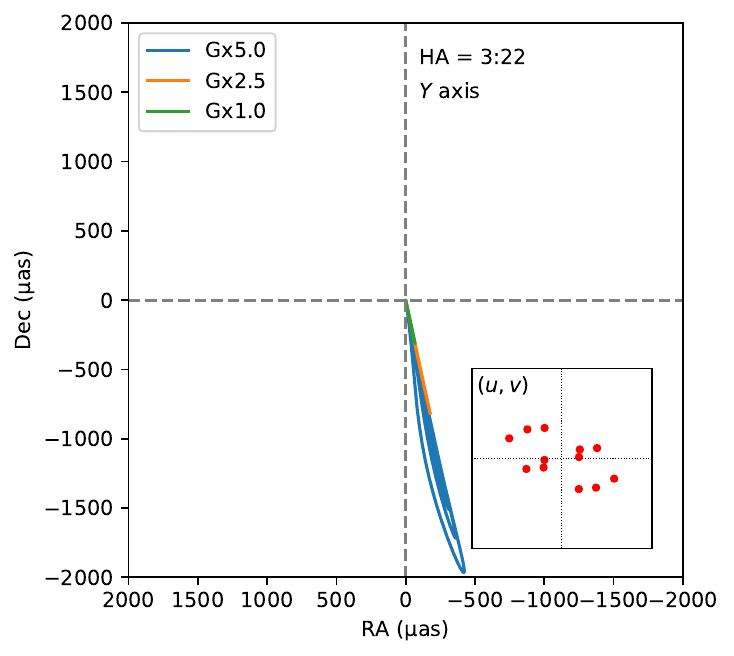} \\
\end{tabular}
\caption{Computation of the astrometric bias with multiples ($\times 1, 2.5, 5$) of a ground level (G) of differential birefringence based on fiber measurements in the case of $\alpha$ Cru and with a degree of linear polarization varying between 0 and 50\% and polarization angles in the $0-2\pi$ range  for different hour angles. The instantaneous $(u,v)$ coverage with the four \redguy{Unit Telescopes} is displayed as an inset in the bottom right of each plot. The full $(u,v)$ coverage is presented in Fig.~\ref{Fig:(u,v)}.}
\label{Fig:bias_alpha_Cru}
\end{figure*}
%
%
%
\subsection{Simulation} 
An example of astrometric bias assuming a partially linearly polarized source and a non-polarizing interferometer is simulated in this section. Aspro2 from the JMMC has been used to simulate the $(u,v)$ points for a fictitious source at the same location as $\alpha$ Cru. The $(u,v)$ tracks are given in Fig.~\ref{Fig:(u,v)}. $(u,v)$ points are selected for three different hour angles labeled with a square, a circle, and a star to compute the astrometric biases. Computations are made using the birefringence measurements of the GRAVITY science channel fibers \citep{Perrin2024a} and organizing the signs to maximize the effect of differential birefringence:
\begin{equation}
\label{eq:GRAVITY_birefringence}
 \left\{\begin{array}{@{}l@{}}
\Delta\psi_{\mathrm{UT_1}}=7.25^\circ, \\
\Delta\psi_{\mathrm{UT_2}}=-13.57^\circ,  \\
\Delta\psi_{\mathrm{UT_3}}=-8.88^\circ,  \\
\Delta\psi_{\mathrm{UT_4}}=12.56^\circ.  \\
\end{array}\right.
\end{equation}
%
%
At this point, we chose a wavelength of $\lambda = 2\,\mu$m. A value of $\frac{\pi}{3}$ is assumed for $\theta$ as it is intermediate between 0 and $\frac{\pi}{2}$ and does not set $\cos 2\theta$ nor $\sin 2\theta$ to 0. The amount of differential birefringence is indicated for the values of Eq.~\ref{eq:GRAVITY_birefringence} (green) and for 2.5 times (orange) and 5 times (blue) that ground level. For each subset, the degree of polarization is varied from 0 to 50\% and the EVPA covers the $0-2\pi$ range. The computed biases are plotted in Fig.~\ref{Fig:bias_alpha_Cru} for the two polarization axes $x$ and $y$ and in combined mode for each of the three selected hour angles. 
The immediate conclusion one can draw from these graphs is that the bias is very much stretched in a single direction as anticipated with the first-order analysis. For moderate differential birefringence, the bias is a line segment. For the lowest level of differential birefringence, a 50\% linear polarization leads to a maximum bias on the order of $100\,\mu$as. The second conclusion is that the effect of differential birefringence combined with source polarization is easily detectable given the characteristic signatures on the $x$ and $y$ axis that makes these biases  simple to calibrate a priori.

\section{Conclusions}
\label{sec:conclusion}
The theoretical expressions of the biases of long-baseline interferometer observables   established in this paper are aimed at interferometers whose polarimetric characteristics can be exactly or approximately described by a diattenuation and a retardance. The bias on the fringe contrast was already well known prior to this sudy. General expressions were derived for the interferograms with split or combined polarization axes for natural, polarized, and partially polarized light (only linear and circular polarization are studied here, but our conclusions can easily be extended to elliptical polarization). These biases were studied for the particular case of symmetric interferometers, as this is how interferometers are built to maximize coherence. 

Theoretical expressions of the bias on the visibility phase have been established for both polarizing and non-polarizing interferometers. It is remarkable that in combined mode for a non-polarizing interferometer, the visibility phase measurements are unbiased for natural light and for partially circularly polarized light. Analytical expressions are given for the other cases. It is shown that for weak diattenuation and retardance, as well as for moderately polarized sources, closure phases are immune to biases to the second order (no matter the type of light). If retardance is achromatic, then the differential phase is also immune to differential retardance. Finally, we investigated the bias on astrometry. It is shown that the bias depends on the $(u,v)$ sampling with an extreme case when the $(u,v)$ points are aligned on a line crossing the origin, in which case the bias can be arbitrarily large. In any other case, the bias can be computed if the retardance of the interferometer is known for each beam. The astrometric bias is expanded for small polarization degrees and small differential retardance. It is shown that in this case, the astrometric bias lies on a straight line crossing the astrometric reference point for non-polarizing interferometers. If the degree of linear polarization varies during the observations, then the astrometric bias has a remarkable signature as it describes a section of the line. If the interferometer is slightly polarizing, then a fixed offset is to be added without changing the general shape of the bias.

%
%
%

%
%

\begin{acknowledgements}
This research has made use of the Jean-Marie Mariotti Center \texttt{Aspro}
service \footnote{Available at http://www.jmmc.fr/aspro}.
\end{acknowledgements}

\bibliographystyle{aa} 
\bibliography{references_pol_bir}

\appendix
\section{Equation details}
\subsection{Derivation of Equation \ref{eq:basic_x}}
\label{sec:eq14_det}
This section describes how Eq.\ref{eq:basic_x} is derived from Eq.~\ref{eq:start}, which is recalled here:
\begin{equation}
\begin{split}
\tilde{I}_{\bullet \mathrm{p},x} = \mathrm{Re}\left[ e^{-i\eta} \left< \right.\right.&  \tau_{x_1}\tau_{x_2}E_x^2\cos\theta_1\cos\theta_2 e^{-i(\psi_{{x_1}}-\psi_{{x_2}})} 
+ \tau_{{y_1}}\tau_{{y_2}}E_y^2\sin\theta_1\sin\theta_2 e^{-i(\psi_{{y_1}}-\psi_{{y_2}})} \\
&\!\!\!\!\!\!\left.\left.- \tau_{x_1}\tau_{{y_2}}E_xE_y\cos\theta_1\sin\theta_2 e^{-i(\varphi_x-\varphi_y+\psi_{{x_1}}-\psi_{{y_2}})} 
- \tau_{x_2}\tau_{{y_1}}E_xE_y\cos\theta_2\sin\theta_1 e^{-i(\varphi_y-\varphi_x+\psi_{{y_1}}-\psi_{{x_2}})} \right>\right].
\end{split}
\end{equation}

Given the notations defined in Eq.~\ref{eq:def}, we can write:
\begin{equation}
\begin{array}{@{}l@{}l@{}}
-(\psi_{x_1} - \psi_{x_2})+\Delta \mathrm{A}\psi = +\frac{1}{2}\Delta\Delta\psi,  \\
-(\psi_{y_1} - \psi_{y_2})+\Delta \mathrm{A}\psi = -\frac{1}{2}\Delta\Delta\psi,  \\
-(\varphi_x-\varphi_y+\psi_{x_1} - \psi_{y_2})+\Delta \mathrm{A}\psi = -\gamma, \\
-(\varphi_y-\varphi_x+\psi_{y_1} - \psi_{x_2})+\Delta \mathrm{A}\psi = +\gamma. \\
%
\end{array}
\end{equation}
Now, using the definition of $\rho$ and $\phi$ (given in Eq.~\ref{eq:rho_phi}), Eq.~\ref{eq:start} becomes%
\begin{equation}
\begin{split}
\tilde{I}_{\bullet \mathrm{p},x} = \mathrm{Re}\left[ e^{-i(\eta+\Delta \mathrm{A}\psi )} \left< \right.\right.&  \rho^2\cos^2\phi\cos\theta_1\cos\theta_2 e^{\frac{i}{2}\Delta\Delta\psi } 
+ \ \rho^2\sin^2\phi\sin\theta_1\sin\theta_2 e^{-\frac{i}{2}\Delta\Delta\psi } \\
&\!\!\!\!\!\!\left.\left.- \tau_{x_1}\tau_{{y_2}}E_xE_y\cos\theta_1\sin\theta_2 e^{-i\gamma} 
- \tau_{x_2}\tau_{{y_1}}E_xE_y\cos\theta_2\sin\theta_1 e^{i\gamma} \right>\right]
\end{split}
.\end{equation}
Using the trigonometric formulae for $\cos\theta_i\cos\theta_j$, $\sin\theta_i\sin\theta_j$ and $\cos\theta_i\sin\theta_j$, we get the result of Eq. \ref{eq:basic_x}:
\begin{equation}
\begin{split}
2\tilde{I}_{\bullet \mathrm{p},x} = \mathrm{Re}\left[ e^{-i(\eta+\Delta \mathrm{A}\psi)} \left< \rho^2 \right. \right.&  \left[ \left[ \cos(\theta_1+\theta_2)\cos2\phi +\cos(\theta_1-\theta_2) \right]\cos(\tfrac{1}{2}\Delta\Delta\psi)+i \left[ \cos(\theta_1+\theta_2)+\cos(\theta_1-\theta_2)\cos2\phi \right]\sin(\tfrac{1}{2}\Delta\Delta\psi)\right]\\
&\!\!\!\!\!\!\!\!\!\!\!\!\!\!\!\!\!\!\!\!\!\!\!\!\!\!\!\!\!\!\!\!\!\!\!\!\!\!\!\!\!\!\!\!\!\!\!\!\!\!\!\!\!\!\!\!\!\!\!\!\!\!\!\!\!\!\!\!\!\!\!\!\!\!\!\!\!\!\!\!\!\!\!\!\left.\left.-E_xE_y \left[ \cos\gamma \left[ (\tau_{x_1}\tau_{{y_2}}+\tau_{x_2}\tau_{{y_1}})\sin(\theta_1+\theta_2) -(\tau_{x_1}\tau_{{y_2}}-\tau_{x_2}\tau_{{y_1}})\sin(\theta_1-\theta_2)\right]
+i\sin\gamma \left[ -(\tau_{x_1}\tau_{{y_2}}-\tau_{x_2}\tau_{{y_1}})\sin(\theta_1+\theta_2) +(\tau_{x_1}\tau_{{y_2}}+\tau_{x_2}\tau_{{y_1}})\sin(\theta_1-\theta_2)\right]\right]\right>\right].
\end{split}
\end{equation}

\subsection{Derivation of Equation \ref{eq:phase_shift}}
This section describes how Eq. \ref{eq:phase_shift} is derived from Eq.~\ref{eq:xi_polarizing}. Weak diattenuation ($\delta \tau = \tau_x - \tau_y$) and weak differential retardance ($\Delta\Delta\psi$) are assumed meaning $\delta\tau$ and $\Delta\Delta\psi$ are considered first-order quantities. A moderate degree of polarization of the source is also assumed meaning $P$ is also considered as a first order quantity. In what follows, \say{$x^{\mathrm{th}}$ order} means \say{$x^{\mathrm{th}}$ order in $\delta\tau$, $\Delta\Delta\psi$ or $P.$} The idea is to derive $\xi$ from $\tan \xi$ by taking the ratio of the two lines of Eq.~\ref{eq:xi_polarizing}. Since $\sin \xi$ is proportional to $\sin(\tfrac{1}{2}\Delta\Delta\psi)$ in this equation, it is at most of order of 1 as well as $\xi$ since  $\sin \xi \simeq \xi$ to first order. As a consequence, the tangent can be expanded to first order as $\tan \xi \simeq \xi$, which indeed is also an expansion to second order since the second order term of the tangent is equal to 0. In addition, to get an expansion of $\xi$ to the second order and since the zeroth order term of $\sin \xi$ is null, the expression of $\cos \xi$ only needs to be expanded to first order.\\ \\
With these assumptions and these considerations, Eq.~\ref{eq:xi_polarizing} can be expanded according to
\begin{equation}
 \left\{\begin{array}{@{}l@{}}
2^{|\varepsilon|}\rho\cos \xi \simeq \tau^2\left[ 1+\varepsilon P (\cos2\theta\cos2\alpha-\sin2\theta\sin2\alpha\cos\gamma) +\frac{\delta\tau}{\tau}\varepsilon\cos2\theta \right], \;\;\;\;\;\;\mathrm{to\;first\;order,}\\
2^{|\varepsilon|}\rho \sin \xi \simeq \frac{1}{4}\tau^2\Delta\Delta\psi\left[ 2\varepsilon\cos2\theta+2\frac{\delta\tau}{\tau}+2P\cos2\alpha   \right], \;\;\;\;\;\;\;\;\;\;\;\;\;\;\;\; \;\;\;\;\;\;\;\;\;\;\;\;\;\;\;\;\;\;\;\,\,\mathrm{to\;second\;order.}
\end{array}\right.
\end{equation}
As a consequence, the expansion of $\tan\xi\simeq\xi$ to second order is equal to
\begin{equation}
\tan\xi\simeq\xi \simeq \frac{1}{4}\Delta\Delta\psi\left[ 2\varepsilon\cos2\theta+2\frac{\delta\tau}{\tau}+2P\cos2\alpha \right] \times \left[ 1-\varepsilon P (\cos2\theta\cos2\alpha-\sin2\theta\sin2\alpha\cos\gamma)-\frac{\delta\tau}{\tau}\varepsilon\cos2\theta \right].
\end{equation}
Keeping second order terms only in the product and grouping terms leads to
\begin{equation}
\xi \simeq \frac{1}{4}\Delta\Delta\psi\left[ 2\varepsilon\cos2\theta+2(P\cos2\alpha+\frac{\delta\tau}{\tau})(1-\varepsilon^2\cos^2\theta)+\varepsilon^2 P \sin4\theta\sin2\alpha\cos\gamma) \right]
.\end{equation}
Hence, the result of Eq. \ref{eq:phase_shift} when applied to baseline $i$, as noted in Sect. 5.  \ %
\vspace{1cm}
\end{document}